\documentclass[12pt,preprint2]{aastex}

\usepackage{epstopdf}
\usepackage{url}

\shorttitle{Galaxy morphology catalog}
\shortauthors{Kuminski \& Shamir}

\begin{document}



\title{Computer-generated visual morphology catalog of $\sim$3,000,000 SDSS galaxies}


\author{Evan Kuminski 
and Lior Shamir$^{*}$ 
}
\affil{Department of Computer Science, Lawrence Technological University, MI 48075 \\ $^{*}$email: lshamir@mtu.edu}




\begin{abstract}
We applied computer analysis to classify the broad morphological type of $\sim3\cdot10^6$ SDSS galaxies. The catalog provides for each galaxy the DR8 object ID, right ascension, declination, and the certainty of the automatic classification to spiral or elliptical. The certainty of the classification allows controlling the accuracy of a subset of galaxies by sacrificing some of the least certain classifications. The accuracy of the catalog was tested using galaxies that were classified by the manually annotated {\it Galaxy Zoo} catalog. The results show that the catalog contains $\sim$900,000 spiral galaxies and $\sim$600,000 elliptical galaxies with classification certainty that has a statistical agreement rate of $\sim$98\% with {\it Galaxy Zoo} debiased ``superclean'' dataset. 
The catalog also shows that in rare cases objects assigned by SDSS pipeline with a relatively high redshift (z$>$0.4) can have clear visual spiral morphology.
The catalog can be downloaded at \url{http://vfacstaff.ltu.edu/lshamir/data/morph_catalog}, and can be accessed through public tables on CAS: public.broadMorph.LargeGM, public.broadMorph.LargeWnnGM, and public.broadMorph.SpectraGM. The image analysis software that was used to create the catalog is also publicly available.



\end{abstract}





\section{Introduction}
\label{introduction}

Autonomous digital sky surveys have been becoming increasingly dominant in astronomy research, and the trend is bound to continue \citep{borne2013virtual,djorgovski2013sky}. Digital sky surveys have powerful image acquisition capabilities, allowing them to collect and store data of very large image databases of astronomical objects. One of the bottlenecks in processing astronomical data is the analysis of the morphology of celestial objects. While substantial information about an astronomical object lays within its morphology and visual structure, automatic morphological analysis requires advanced computer vision algorithms, making the analysis of millions of astronomical objects a challenge.

Several methods for automatic galaxy image analysis have been proposed, including GALFIT \citep{pen02}, GIM2D \citep{sim99}, CAS \citep{con03}, Gini \citep{Abr03}, Ganalyzer \citep{sha11}, and SpArcFiRe \citep{davis2014sparcfire}, as well as image analysis methods based on machine learning \citep{sha09,huertas2009robust,banerji2010,kum14,dieleman2015rotation}.

One of the goals of these methods is to analyze the morphology of galaxies in large galaxy image databases, so that the morphological information can be part of data releases and used similarly to the other information collected by digital sky surveys. However, artifacts, imperfect images, limited resolution, saturated objects, small object surface size, and misidentified objects introduce a challenge when applying these methods to produce clean automatically generated catalogs of galaxy morphology.

Manual inspection of galaxy images can provide accurate annotation, but due to the labor-intensive nature of the task such catalogs can practically contain several thousand galaxies \citep{efigi,nair2010catalog}. To allow the analysis of larger databases, manual annotation was carried out by non-astronomer volunteers, accessing the images and submitting their annotation via the internet \citep{lintott2008galaxy} to generate catalogs of several hundred thousands annotated galaxies such as Galaxy Zoo 1 \citep{galaxyzoo1} and Galaxy Zoo 2 \citep{galaxyzoo2}. However, even when using the power of citizen scientists the annotation process is still limited by the amount of data it can analyze, requiring several years and substantial labor to produce catalogs such as Galaxy Zoo. That bandwidth problem will be magnified when more powerful sky surveys such as the Large Synoptic Sky Survey \citep{lsst} see first light, and acquire billions of galaxy images.

Using non-professional astronomers as data annotators, without the ability to control their skill, experience, and quality of their individual annotations, clearly adds noise into the system. Although the presence of malicious annotators is mild \citep{galaxyzoo1}, participants who are not trained scientists might make wrong or careless annotations, making the data unclean. To correct for human error it is required to apply statistical methods that ignore objects on which the voting was not consistent across the data annotators. In Galaxy Zoo, a classification of a galaxy is defined ``clean'' if 80\% or more of the voters who classified that galaxy provided the same answer \citep{galaxyzoo1}. A classification of a galaxy is defined ``superclean'' if it has an agreement of 95\% or more of the votes. Filtering classifications that did not meet a certain level of agreement threshold can provide a cleaner subset of the data, but at the cost of sacrificing some of the samples that their voting did not meet the ``clean'' or ``superclean'' criteria. For instance, for question 1 (`'Is the galaxy simply smooth and rounded, with no sign of a disk?'') in Galaxy Zoo 2, just $\sim$7.8\% of the galaxies that were classified in the entire dataset met the ``superclean'' criterion \citep{kum14}. That is, an average number of 44 citizen scientists \citep{galaxyzoo2} voted on each galaxy to indicate whether it is smooth and rounded or not, but different citizen scientists provided different answers on the same question and the same galaxy. Only in $\sim$7.8\% of the galaxies that were classified the agreement among the voters was 95\% or higher. That shows that even when using a very high number of data annotators not all celestial objects can be classified with conclusive results.

The inability of manual annotation to perform exhaustive analysis of extremely large databases reinforces the need for automatically generated galaxy morphology catalogs. \cite{sim11} performed and released data from automatic analysis of Sersic and disk models of over $10^6$ SDSS galaxies. \cite{Hue10} released an automatically generated catalog of Sloan Digital Sky Survey  galaxies with spectra, and classified them into four basic Hubble morphological types of E, S0, Sab, and Scd. That work was followed by an automatically generated catalog of visual morphologies of $\sim$50,000 CANDELS galaxies \citep{huertas2015catalog,huertas2015morphologies}.

Here we applied computer vision and pattern recognition methodology to analyze the broad morphology of $\sim$3 million galaxies taken from Sloan Digital Sky Survey (SDSS) DR8. The catalog is publicly available, as well as the source code of the method that was used to analyze the galaxy images. The methodology can be used to analyze the broad morphology of existing and future larger databases of galaxy images such as the Large Synoptic Survey Telescope.

\section{Methods}
\label{methods}


The initial set of galaxies was selected from SDSS \citep{sdss} DR8 through the Catalog Archive Server (CAS) based on the constraints described in Table~\ref{query}.

\begin{table*}[ht]
\begin{center}
\caption{Attributes of the CAS query used to select the initial set of galaxies}
\label{query}
\begin{tabular}{cll}
\tableline\tableline
Field & Constraint & Description \\
\hline
type & =3 & Objects identified as galaxies by SDSS pipeline \\
petroRad\_r & $>$5.5 & Petrosian radius greater than 5.5 arcseconds \\
petroRadErr\_r & $<$5 & Petrosian radius error smaller than 5 arcseconds \\
flags \& 0x0000000001000000 & =0 & The object is not defined as ``too large'' to be \\
                                             &     &  measured by SDSS pipeline \\

flags \& 0x0000000000008000 & =0 & The object is not defined as ``bad radial'' by SDSS pipeline \\
flags \& 0x0000000000000400 & =0 & The Petrosian radius is not beyond the last point in \\
                                             &     &  the radial profile \\

flags \& 0x0000000000400000 & =0 & The sky is not defined as ``bad sky'' by SDSS pipeline \\
flags \& 0x0000000000000004 & =0 & The object is not too close to the edge of the frame \\
flags \& 0x0000002000000000 & =0 & Peaks were not too close \\
flags \& 0x0000000000000200 & =0 & Only one Petrosian radius was found \\
flags \& 0x0001000000000000 & =0 & No child of the object had too few good detection \\
\tableline
\tableline
\end{tabular}
\end{center}
\end{table*}

The query aimed at selecting just galaxies of a sufficiently large size that allows the analysis of their morphology, and avoided as much as possible saturated images or images with multiple peaks or Petrosian radii that could be galaxy mergers or multiple sources that make the morphology of the object more complex. 

The vast majority of the objects returned by the query did not have spectra. However, many of the objects with spectra are small or faint, and therefore often do not allow good visual identification of the morphology. Larger objects are likely to lead to a higher identification rate, as larger objects contain more noticeable visual cues that can be used for its correct morphological classification.

The query returned 2,939,891 objects, and the SDSS image of each object was downloaded automatically through CAS. The JPG images were saved in the color TIF (Tagged Image File) format, and each image had the dimensionality of 125$\times$125 pixels. To avoid pressure on CAS server, only one image was downloaded at a time. Downloading the images of all objects from CAS was completed after $\sim$25 days.


To handle some of the larger objects that their size exceeded the size of the image, for each image that was downloaded the pixels on the edge of the image that had grayscale value greater than 125 were counted. The initial scale of the image is 0.1 arcseconds per pixel. If more than 25\% of the edge pixels have grayscale value  greater than 125 it is assumed that the edges are not background sky, and therefore the object is too large to fit inside the image. In that case the scale was changed by 0.01 arcseconds per pixel, and the image was re-downloaded and tested again until the object fitted the frame. The 25\% threshold is used since in some cases parts of the galaxy or background objects such as stars can be located on the edge of the image, but since these objects are expected to be small no more than 25\% of the edge pixels are expected to be bright. The simple method ensured that the entire object was inside the frame, surrounded by background sky. 

Although all images that were downloaded were images of objects flagged as galaxies by SDSS, some of the images contained artifacts, errors, or were actually images of astronomical objects that were not galaxies as shown in Table~\ref{artifacts}. Images of these non-galaxy objects can be identified by the image size, as the artifacts, saturated pixels, and error messages changed the compression factor of the image to result in a much larger or much smaller file than a ``clean'' galaxy image. In some cases the image failed to download due to a server error, and since the error message is typed on the image the file becomes larger. By observing 200 images and their file sizes, we set a minimum threshold of 2000 bytes and an upper threshold of 6000 bytes. Based on our observations, objects with file size outside that range were not actual astronomical objects. Therefore, these images were removed from the catalog, resulting in the sacrifice of 27,992 images. Table~\ref{artifacts} shows several examples of images and their file sizes.

\begin{table}[h!]
\begin{center}
\caption{Galaxy images and non-galaxy images in SDSS. All images were flagged by SDSS as galaxies, although some are clearly artifacts. The file size provides a simple mechanism to remove images of artifacts from the dataset}
\label{artifacts}
\begin{tabular}{cll}
\tableline\tableline
SDSS  DR8 ID & File size & Image \\
                     &  (KB)     & \\
\hline
1237645879037067801 & 3.13 & \includegraphics[angle=00,scale=.50]{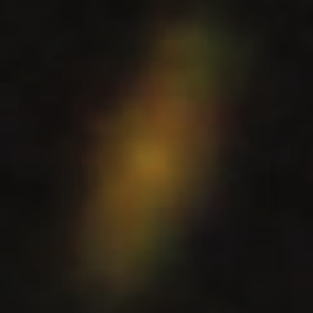} \\
1237649743943369548 & 3.34 & \includegraphics[angle=00,scale=.50]{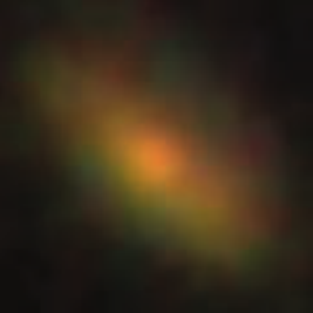} \\
1237671261735354668 & 3.47 & \includegraphics[angle=00,scale=.50]{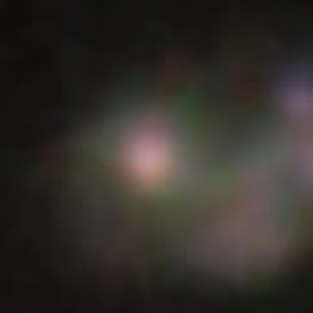} \\
1237646012176400498 & 16.4 & \includegraphics[angle=00,scale=.50]{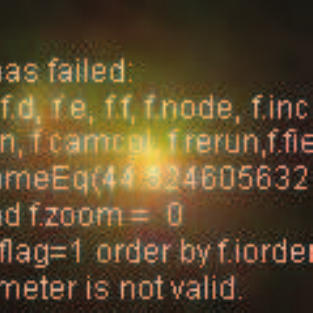} \\
1237645941290369211 & 8.96 & \includegraphics[angle=00,scale=.50]{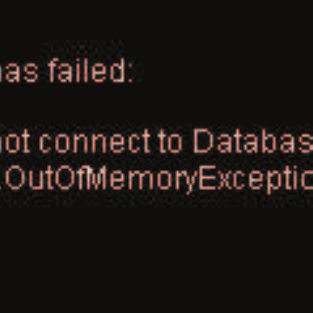} \\
1237649769179578519 & 11.3 & \includegraphics[angle=00,scale=.50]{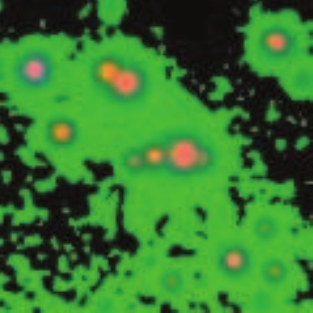} \\
1237672815434465362 & 8.5 & \includegraphics[angle=00,scale=.50]{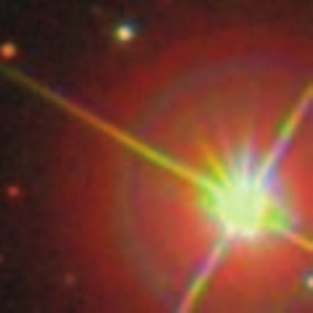} \\
1237651752387411988 & 1.57 & \includegraphics[angle=00,scale=.50]{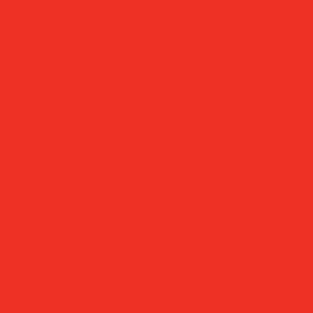} \\
\tableline
\tableline
\end{tabular}
\end{center}
\end{table}

\subsection{Galaxy image analysis}
\label{method}


The image analysis method used to classify the images is Wndchrm \citep{Sha08,shamir2010impressionism,shamir2013wnd}, that first computes 2885 numerical descriptors from each image such as textures \citep{haralick1973textural,tamura1978textural}, edges \citep{prewitt1970object}, shapes \citep{orlov2008wnd}, statistical distribution of the pixel intensities \citep{hadjidemetriou2001spatial}, polynomial decomposition of the image \citep{teague1980image}, and fractal features \citep{wu92}. These features are extracted from the raw pixels, as well as the image transforms and multi-order image transforms \citep{shamir2010impressionism}. A complete and detailed description of the set of numerical image content descriptors and comprehensive performance analysis is available in \citep{Sha08,orlov2008wnd,shamir2010impressionism,shamir2009knee}. The source code is also publicly available \citep{shamir2013wnd}, and can be downloaded at \url{http://vfacstaff.ltu.edu/lshamir/downloads/ImageClassifier}.

In particular, Wndchrm has been proven to be informative for analysis of galaxy morphology, and was found useful for tasks such as galaxy classification \citep{sha09, kum14}, unsupervised analysis of galaxy images \citep{schutter2015galaxy,shamir2013automatic}, and automatic detection of peculiar galaxies \citep{shamir2012automatic,shamir2014automatic,shamir2014chloe}.

Previous experiments have shown that Wndchrm achieves $\sim$95\% of classification accuracy when separating spiral and elliptical galaxies \citep{sha09}, which outperforms methods such as Gini \citep{Abr03} and CAS \citep{con03} as described in \citep{sha09}. Its performance is favorably comparable to the performance reported by using a Neural Network classifier \citep{banerji2010}, and is also higher than methods based on the data provided by SDSS photometry pipeline \citep{yamauchi2005morphological}. The accuracy can also be improved when ignoring classifications with lower certainty \citep{sha09}. A detailed description and experimental results of Wndchrm classification to spiral and elliptical galaxies is available in \citep{sha09}.

The training set with which the classifier was trained included 150 galaxies classified manually as spiral galaxies, and 150 galaxies classified manually as elliptical galaxies as was done in \citep{sha09}. The training images were first selected by manually choosing 80 galaxy images from each morphological type, and then using a classifier trained with these samples to classify a dataset of 1000 galaxy images. The classified images were inspected manually, and when a misclassified galaxy was noticed the process was stopped, the misclassified galaxy was added to the correct class in the training set, and the classifier was re-trained using the new training set. The process was repeated until no misclassified galaxies were observed.

\begin{figure*}[ht]
\includegraphics[scale=0.45]{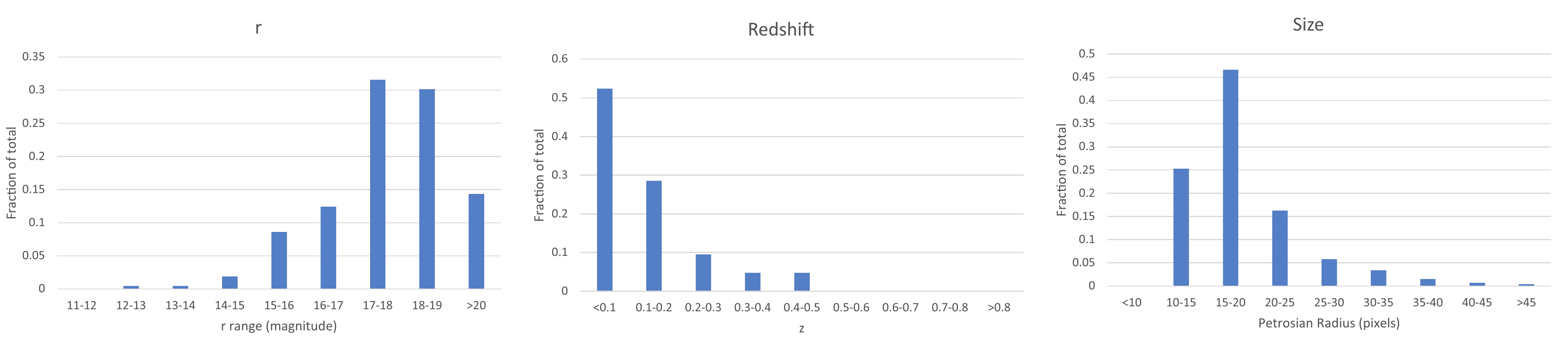}
\caption{Histograms of the r magnitude, Petrosian radius, and photometric redshift of the training galaxies.}
\label{histogram_training}
\end{figure*}

The distribution of the Petrosian radius measured on the r band, r magnitude, and photometric redshift of the training galaxies are displayed in Figure~\ref{histogram_training}.

As a classifier, Wndchrm does not just predict the class of a given sample, but also provides the user with the certainty of the classification to each class as a value within the interval [0,1] \citep{Sha08,shamir2010impressionism}. For instance, certainty of 1 means a high certainty of the classification of the object to that class, while if the certainty of the classification is 0.7 there is a higher chance that the classification is incorrect. The sum of certainties of a sample to belong in all classes is always 1.0 \citep{Sha08,sha09}. These computed certainties can be used in a similar fashion to the degree of agreement between the voters in a citizen science campaign. Therefore, thresholding these certainty values can be used to provide subset catalogs with different levels of accuracies, such that galaxies that do not meet the threshold can be considered ``unknown''. That practice allows controlling the accuracy/size trade-off, as will be discussed in Section~\ref{evaluation}.

\section{Catalog of broad galaxy morphology}
\label{catalog}

The catalog of galaxies is available for download at \url{http://vfacstaff.ltu.edu/lshamir/data/morph_catalog} in the form of a CSV (comma separated values) file. The catalog provides information about 2,911,899 galaxies. The information for each object includes the DR8 ID, right ascension, declination, certainty of the galaxy being elliptical, and the certainty of the galaxy being spiral as computed by the image analysis method described in Section~\ref{method}.

The DR8 ID can be used as a handle to extract all other photometric or other information of interest about the galaxies through CAS.

\begin{figure*}[ht]
\includegraphics[scale=0.45]{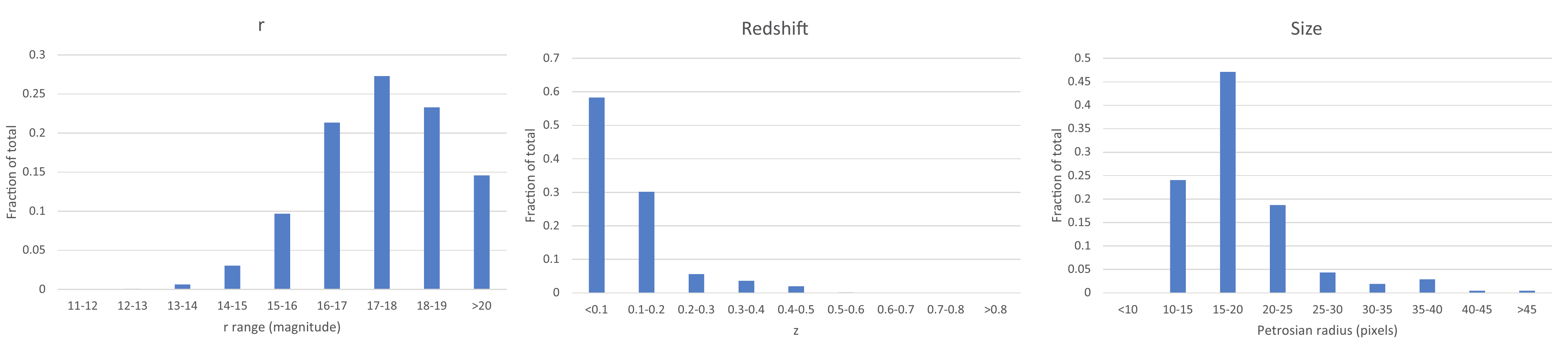}
\caption{Histograms of the r magnitude, Petrosian radius, and redshift of the galaxies in the catalog.}
\label{histogram_catalog}
\end{figure*}

The distribution of the size, r magnitude, and redshift (of objects with spectra) of the training galaxies are displayed in Figure~\ref{histogram_catalog}.

\subsection{Morphological catalog of SDSS objects with spectra}
\label{spectroscopy}

In a similar way we also compiled a catalog of all objects with spectra in DR8. That catalog contains for each object the specObjID, right ascension, declination, z, z error, certainty of classification to elliptical, certainty of classification to spiral, and certainty of classification to a star. 

The catalog includes the morphological analysis of 2,638,823 astronomical objects with spectra. Unlike the catalog described in Section~\ref{catalog}, the objects were not filtered by their size, so the catalog also contains the classification of small or faint objects that their SDSS image does not provide sufficient information to make a clear classification of the morphology.

\begin{figure*}[ht]
\includegraphics[scale=0.45]{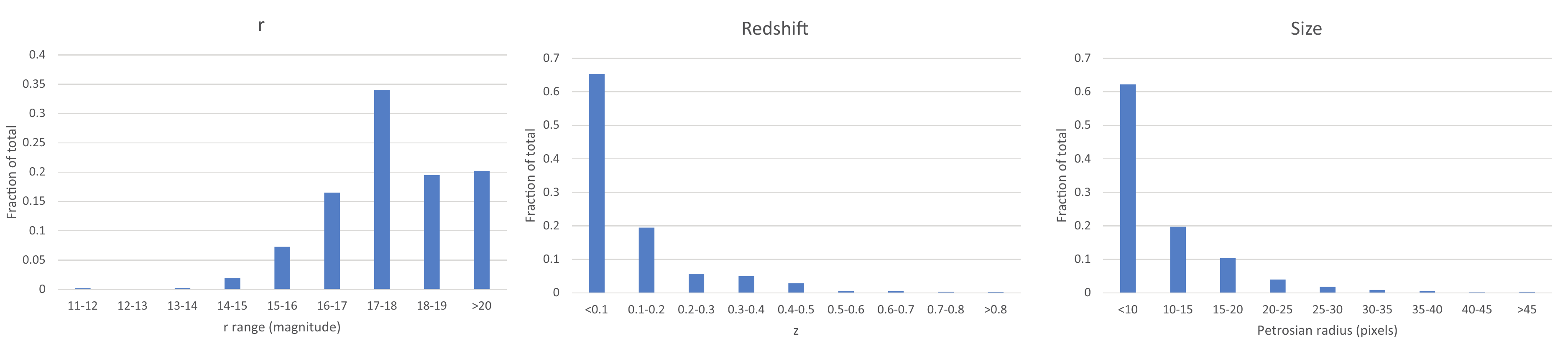}
\caption{Histograms of the r magnitude, size, and redshift of the objects in the catalog.}
\label{histogram_spec_catalog}
\end{figure*}

The distribution of the size, r magnitude, and redshift of the training galaxies are displayed in Figure~\ref{histogram_spec_catalog}.

Because the catalog also includes many objects that are not galaxies, it contains a much larger number of smaller objects. SDSS photometry pipeline provides an automatic separation between stars and galaxies by the difference between the cmodel and PSF magnitude, but that separation is not perfect \citep{stoughton2002sloan}. By analyzing all objects with spectra, the catalog includes objects that could have been misclassified as stars, as well as providing information about objects that were not classified by SDSS photometry pipeline as galaxies. The classification in this catalog is based on the visual cues, and can be used in combination with SDSS PhotObjAll.

\section{Evaluation}
\label{evaluation}

To evaluate the accuracy of the catalog we compared the classifications of the galaxies in the catalog to the manual classifications of the galaxies in the Galaxy Zoo catalog \citep{galaxyzoo1}. As a crowdsouring-based project it is possible that the classifications of the citizen scientists might in some cases disagree with the morphology of the object determined by a professional astronomer. However, the Galaxy Zoo catalog is based on manual classification of a large number of citizen scientists, applies bias correction mechanisms, and has been inspected for its accuracy \citep{galaxyzoo1}, making it suitable to provide a ground truth baseline to test the accuracy of computer-generated catalogs. Although Galaxy Zoo can theoretically contain errors, the number of misclassifications in the debiased ``superclean''  dataset is expected to be low. \textbf{The Galaxy Zoo debiased dataset is corrected for the redshift effect, and therefore the probability of a galaxy to be manually classified as elliptical is not expected to increase as the redshift gets higher}.

Since the number of galaxies analyzed for this catalog is far higher than the number of galaxies in Galaxy Zoo, it is clear that not all galaxies in the catalog are also analyzed by Galaxy Zoo. However, given the large number of galaxies in both datasets, even a mild overlap can provide a sufficient number of galaxies to allow the analysis of the consistency of the classifications across the two datasets. 

The total number of galaxies in the catalog that were also classified by Galaxy Zoo is 229,271. That includes all galaxies that were classified, and clearly some of these galaxies were not classified in high certainty that makes them suitable to be used as ground truth. Therefore, we used the Galaxy Zoo galaxies that were classified as ``superclean'', and were corrected for red-shift bias \citep{galaxyzoo1}. The number of galaxies in the catalog that were also classified by Galaxy Zoo with debiased ``superclean'' classification is 45,377.

\begin{figure}[ht]
\includegraphics[scale=0.65]{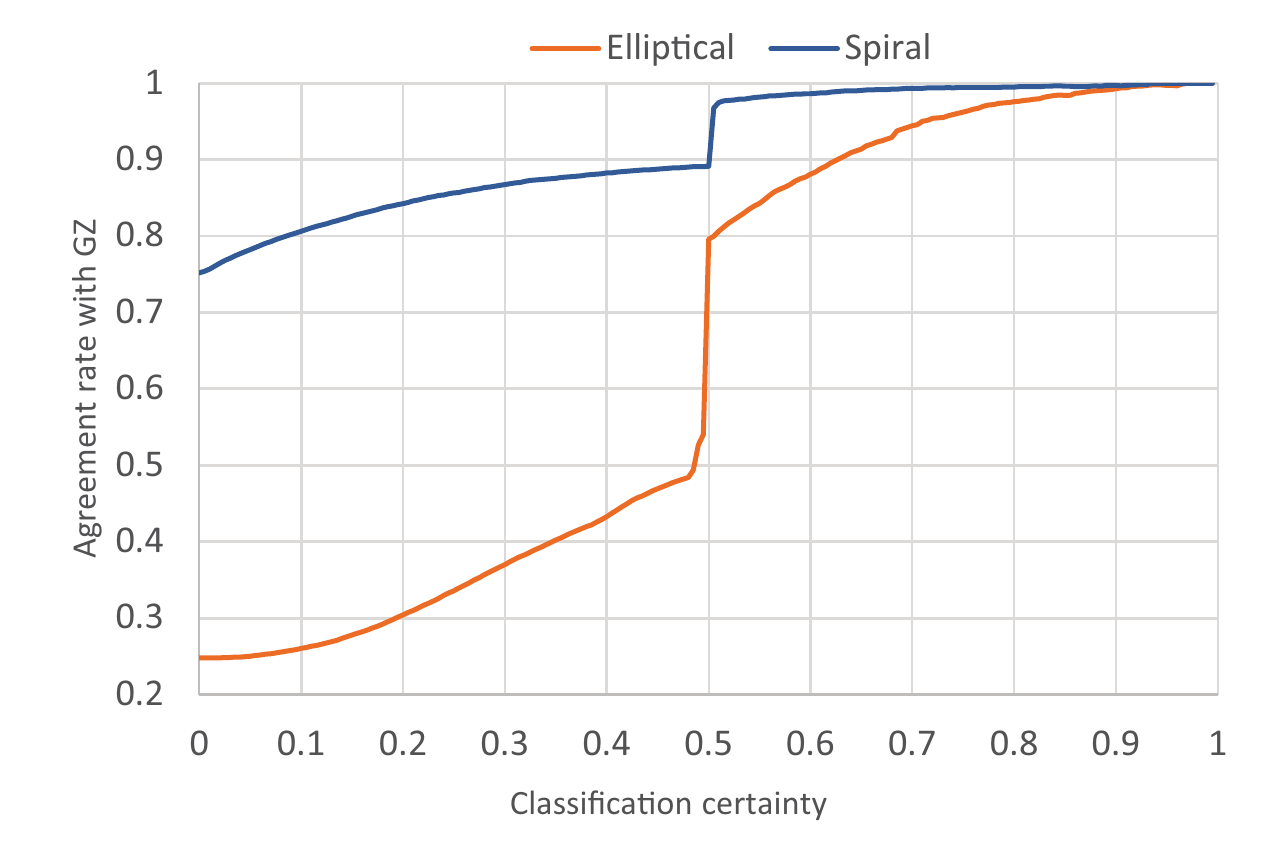}
\caption{Rate of agreement with galaxies classified by Galaxy Zoo ``superclean'' debiased dataset for different ranges of computed classification certainties}
\label{superclean_spiral_elliptical}
\end{figure}

Figure~\ref{superclean_spiral_elliptical} shows the agreement between the computer classification and the Galaxy Zoo ``superclean'' manual classification for different levels of computer classification certainty.
 
As the figure shows, when the computer classification is a spiral galaxy with certainty higher than 0.6, the rate of agreement between the catalog and Galaxy Zoo superclean debiased data is $\sim$98.7\%. That is, of all galaxies classified as spiral with certainty higher than 0.6 and were also classified by Galaxy Zoo superclean dataset, $\sim$98.7\% were also classified as spiral in Galaxy Zoo superclean catalog. 

When the certainty of a spiral galaxy classification is higher than 0.625 the agreement with Galaxy Zoo superclean data is $\sim$99\%, and the rate of agreement is $\sim$98\% when the certainty level is 0.54 or higher. These results show that when setting the certainty threshold to 0.54, the accuracy of the classification is very similar to the classifications of Galaxy Zoo superclean data. 

Similarly, when the classification is elliptical galaxy in certainty of 0.8 or higher, $\sim$97.7\% of the same galaxies that are also classified by Galaxy Zoo superclean dataset are also classified as elliptical. Expectedly, the accuracy leaps when the threshold reaches 0.5, which is the threshold beyond which a galaxy is considered to be more likely to be spiral (when spiral certainty is greater than 0.5) or elliptical (when elliptical classification certainty is greater than 0.5).





The figure also shows that the disagreement rate is higher for the elliptical galaxies compared to the spiral galaxies. That difference can be explained by the higher sensitivity of the human eye to spiral galaxies compared to elliptical galaxies \citep{dojcsak2014quantitative}. That is, when the spiral features of the galaxy are visible it is easier to classify the galaxy as spiral by manual observation. However, when the spirality of the galaxy is not easily noticeable it can mean that the galaxy is indeed not spiral, or that the image resolution is not high enough to identify the spirality. Therefore, it is expected that more galaxies classified as elliptical are actually spiral galaxies, compared to galaxies classified as spiral that are actually elliptical \citep{dojcsak2014quantitative}.

\begin{figure}[ht]
\includegraphics[scale=0.65]{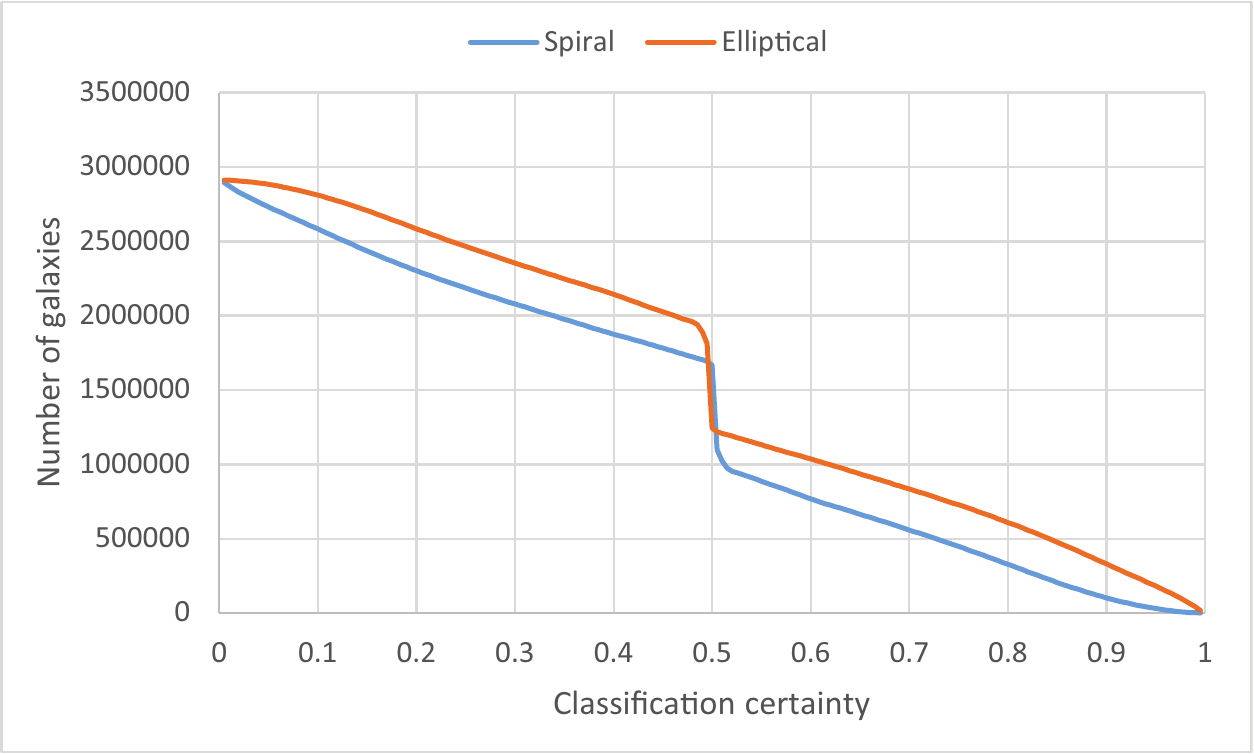}
\caption{Number of spiral and elliptical galaxies classified above different certainty levels.}
\label{counts}
\end{figure}

Figure~\ref{counts} shows the number of galaxies in the catalog classified as spiral or elliptical above different levels of certainty. Expectedly, when the certainty threshold is zero all galaxies meet the threshold, and the number of galaxies that meet the threshold gets smaller as the threshold increases. At certainty level 0.5 a sharp drop is observed, which corresponds to the drop in Figure~\ref{superclean_spiral_elliptical}.

Agreement rate of 98\% with Galaxy Zoo ``superclean'' catalog is achieved when the certainty level for spiral galaxy classification is 0.54 or higher, and the certainty level for elliptical galaxies is 0.8 or higher. Therefore, the catalog contains $\sim$600,000 elliptical galaxies and $\sim$900,000 spiral galaxies that have 98\% accuracy when considering Galaxy Zoo ``superclean'' data as ground truth. Since the computer analysis works differently from the human brain, the galaxies classified manually as ``superclean'' are not necessarily easier to classify by the computer compared to other galaxies that are not ``superclean''. The comparison between human certainty and computer certainty is described in Section~\ref{correlation}.


\subsection{Evaluation of the catalog of objects with spectra}
\label{spec_eval}


Similarly, we also evaluated the accuracy of the catalog of objects with spectra by using Galaxy Zoo debiased ``superclean'' data as ground truth. The number of objects in the catalog that were also classified by Galaxy Zoo is 620,529. Among these galaxies, 59,984 were classified with ``superclean'' accuracy. The high overlap between the objects with spectra and Galaxy Zoo is because Galaxy Zoo intentionally selected objects classified as galaxies by SDSS pipeline and also had spectroscopy information. Objects identified as stars, on the other hand, were not selected for Galaxy Zoo classification since Galaxy Zoo focused on the morphology on galaxies, and therefore asking the citizen scientists to classify the morphology of objects identified by SDSS as stars would be an inefficient use of resources.

\begin{figure}[ht]
\includegraphics[scale=0.65]{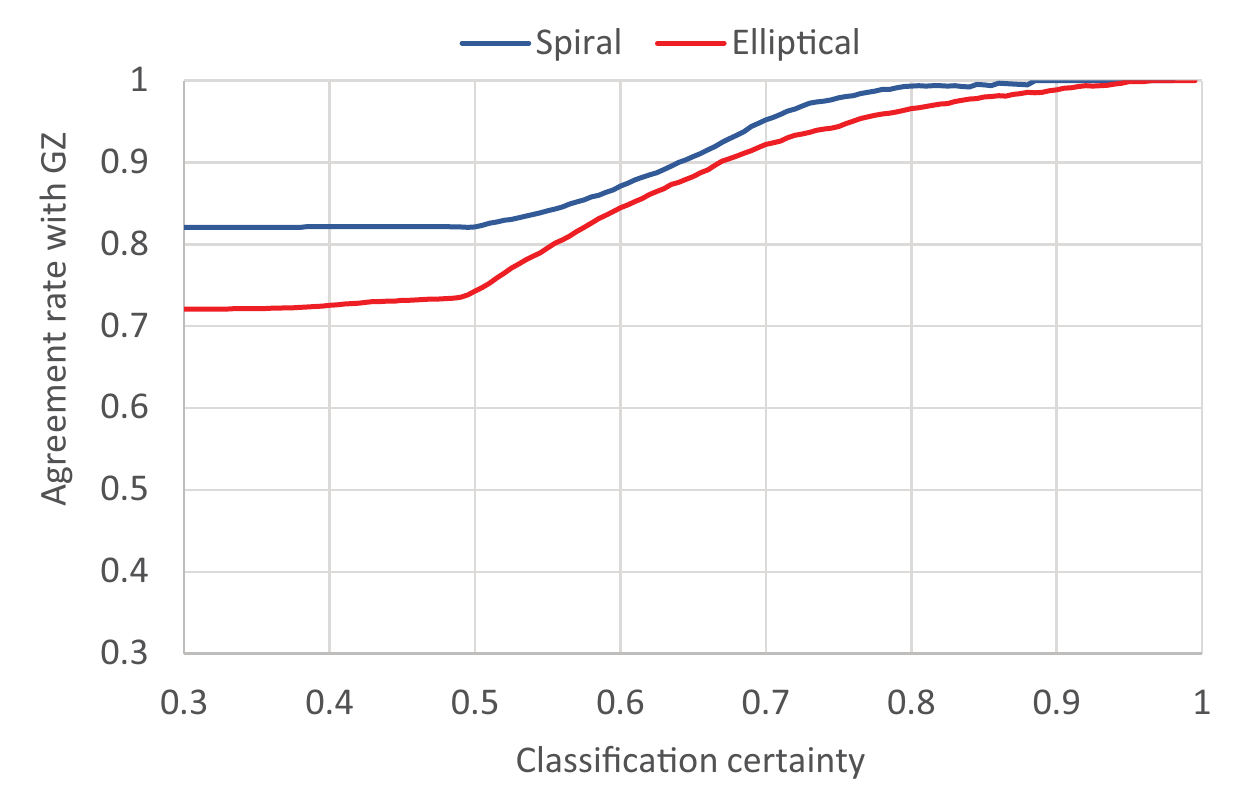}
\caption{Rate of agreement between the catalog of objects with spectra and Galaxy Zoo ``superclean'' debiased dataset}
\label{superclean_spec}
\end{figure}

Figure~\ref{superclean_spec} shows the agreement between the Galaxy Zoo classification of these galaxies and the computer classification for different levels of computer classification certainty. Since the catalog of objects with spectra also contains small or faint objects with morphology that is more difficult to identify, the agreement between the two catalogs is lower compared to the catalog of objects without spectra.

\begin{figure}[ht]
\includegraphics[scale=0.65]{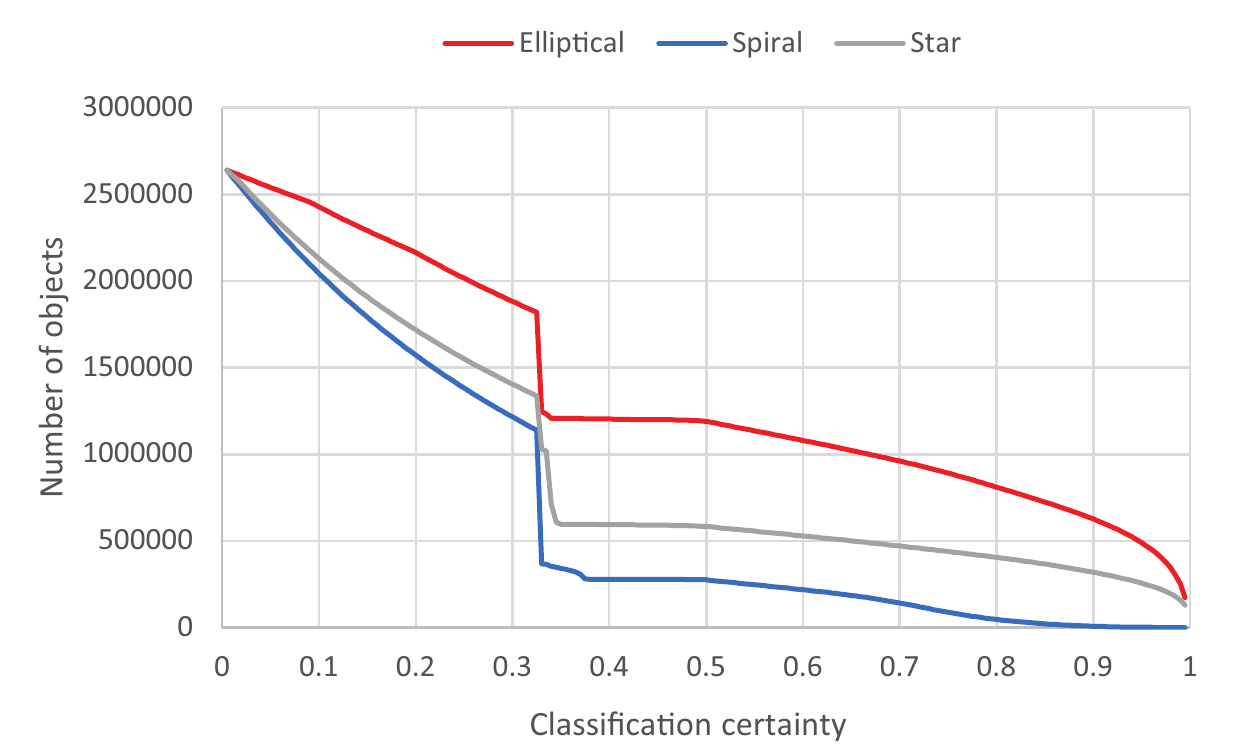}
\caption{Number of spiral galaxies, elliptical galaxies, and stars with spectra classified above different certainty levels.}
\label{counts_spec}
\end{figure}

The number of galaxies classified as spiral with different thresholds of classification certainty are displayed in Figure~\ref{counts_spec}. As the figure shows, the number of objects identified as spiral galaxies is much lower in the catalog of objects with spectra. That can be explained by the fact that the objects in that catalog were not filtered by their size, so small objects with no apparent spirality were more common in that catalog. 

\begin{figure}[ht]
\includegraphics[scale=0.65]{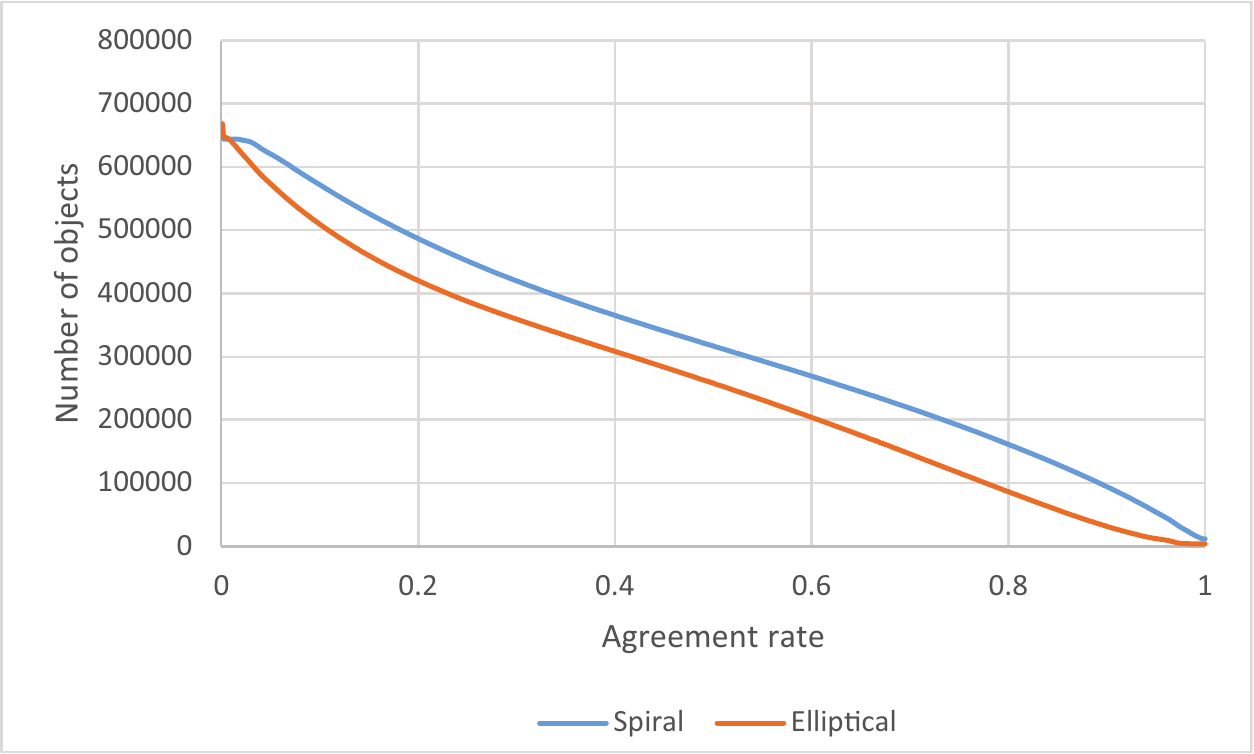}
\caption{Number of spiral and elliptical galaxies with spectra above different agreement rates in Galaxy Zoo debiased dataset}
\label{galaxy_zoo_agreement}
\end{figure}

Figure~\ref{galaxy_zoo_agreement} shows the number of Galaxy Zoo galaxies above different agreement rates. Expectedly, a higher agreement rate leads to fewer galaxies, but the decline is more consistent compared to the computer classification. The decline in the number of galaxies that meets the agreement thresholds starts from agreement rate of 0 because the classifications are not compared to external ground truth, and therefore majority of the votes does not lead to higher consistency with a certain standard of correct classifications.

\section{Correlation between the computer and manual classification}
\label{correlation}

\begin{figure}[ht]
\includegraphics[scale=0.55]{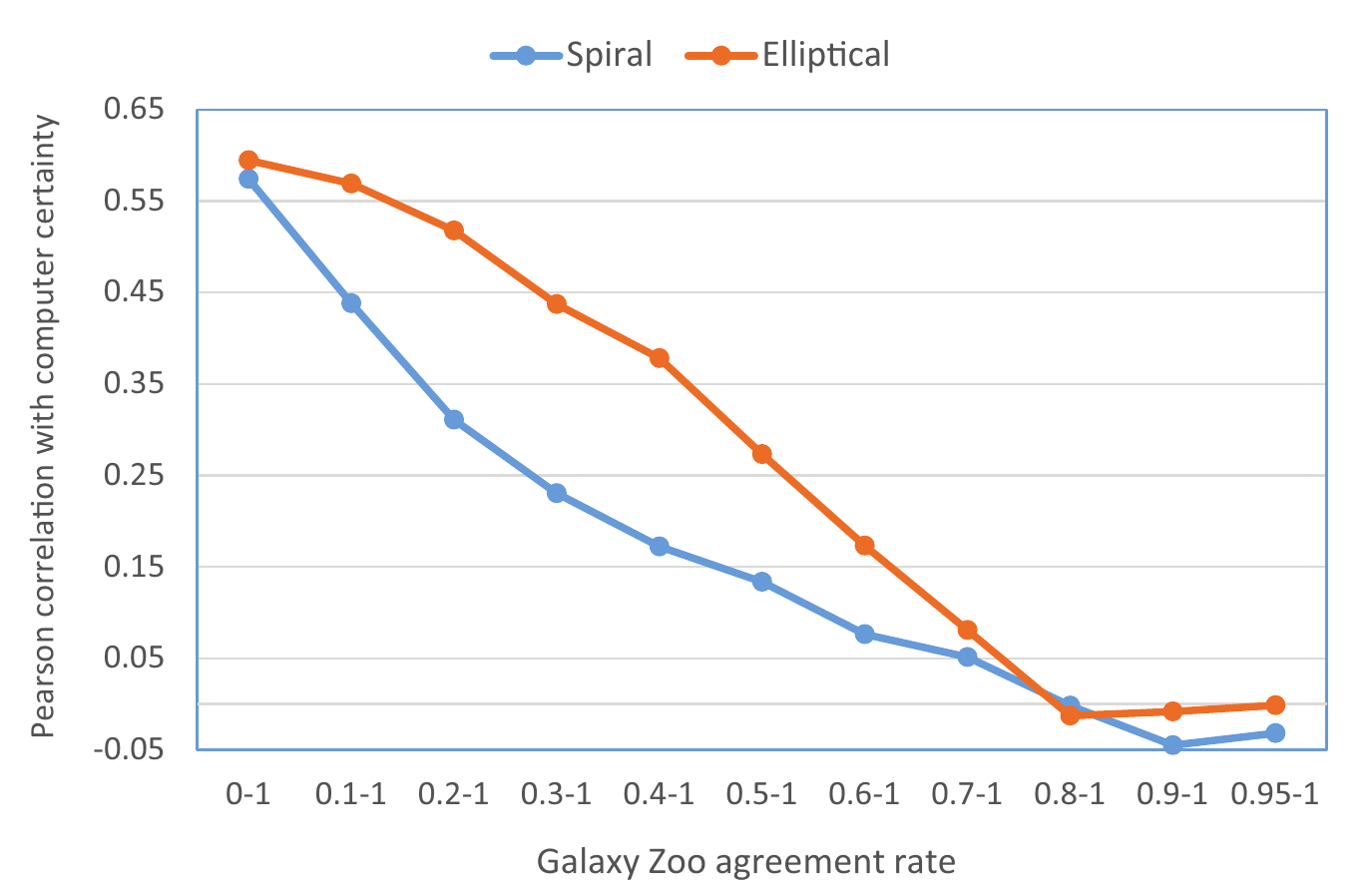}
\caption{Pearson correlation between the manual and computer certainty for different ranges of GZ agreement rate.}
\label{correlation_with_gz}
\end{figure}

When comparing machine and human classification of galaxies, more accurate human classification leads to higher consistency with machine learning methods performing the same task \citep{THMS}. However, the computer analysis is clearly not an attempt to mimic the cognitive processes activated inside the human brain, but merely an attempt to produce similar output as the output that would have been provided by careful manual classification of the images. Therefore, it is not clear whether galaxies that are easy to classify manually are also easier to classify by the computer. Figure~\ref{correlation_with_gz} shows the Pearson correlation between the galaxy zoo agreement rate and the certainty of the computer classification of the same galaxies.

The graph shows that for galaxies with high agreement rate the correlation between the citizen scientist agreement rate, which can be used as an indication for the easiness of the manual classification, has merely weak correlation with the certainty of the classification computed by the machine learning method. The correlation becomes higher when the agreement threshold gets lower, and that can be explained by the fact that the classification of most galaxies in the two catalogs agree, leading to a correlation between the classifications.

It also shows that the correlation gets weaker when the galaxies are classified with higher agreement by the galaxy zoo participants, showing that galaxies that are easier to classify manually are not necessarily classified by the computer with higher certainty. In fact, for galaxies classified by the participants with agreement rate of 80\% or higher there is virtually no correlation with the certainty of the classification computed by the machine learning method.

\begin{figure}[ht]
\includegraphics[scale=0.65]{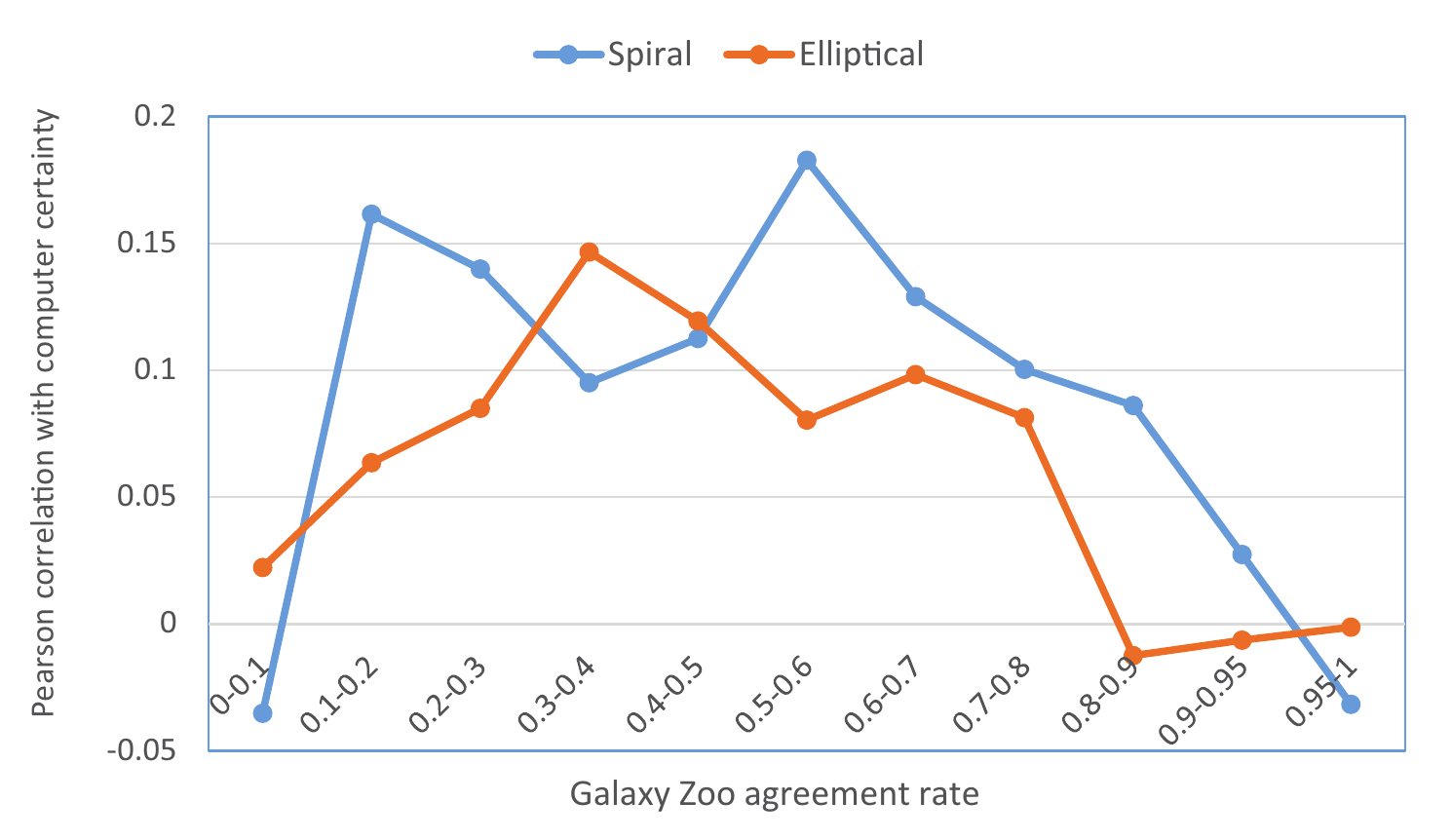}
\caption{Pearson correlation between the manual and computer certainty for different ranges of GZ agreement rates.}
\label{correlation_with_gz_ranges}
\end{figure}

Figure~\ref{correlation_with_gz_ranges} shows the same analysis such that all ranges are equal, so that no bias caused by the broader ranges of the galaxies with lower agreement rate can affect the computed correlation. As in Figure~\ref{correlation_with_gz}, the correlation with the computer certainty gets lower when the galaxies are easier to classify manually.

\begin{figure}[ht]
\includegraphics[scale=0.65]{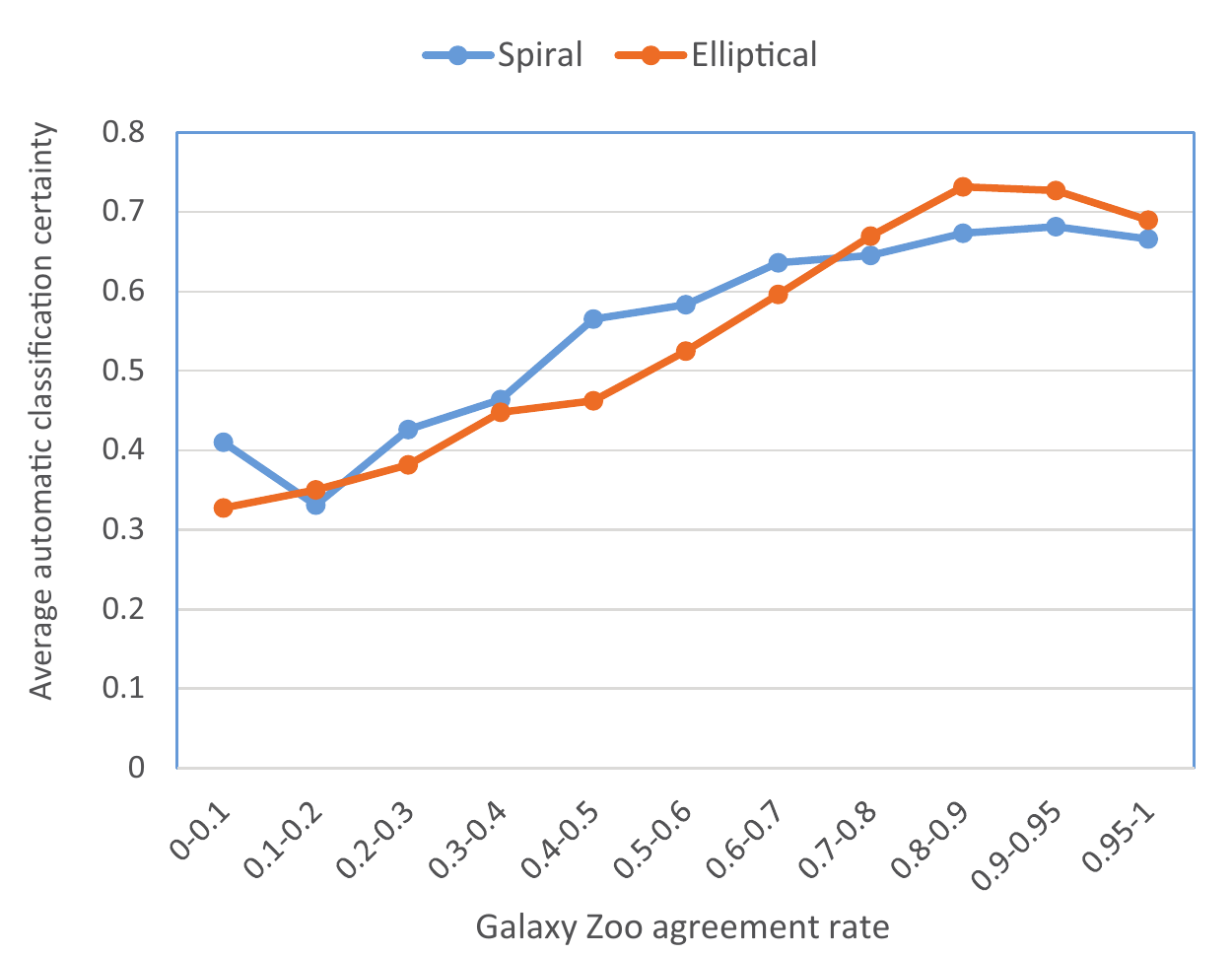}
\caption{Average computer classification certainty in different ranges of GZ agreement rates.}
\label{mean_certainty}
\end{figure}

The observation that galaxies that are easier to classify manually are not necessarily easier to classify automatically is also reflected by the average certainties of the automatic classifications. Figure~\ref{mean_certainty} shows the average certainty of the computer classifications for different ranges of citizen scientist agreement rates.

As the figure shows, when the agreement rate of the citizen scientists is higher than 80\% the average certainty of the automatic classification  of the galaxies does not increase with the agreement rate of the citizen scientists.

\section{Completeness of the catalog}
\label{completeness}

As discussed in Sections~\ref{evaluation}, achieving a higher classification accuracy requires the sacrifice of many of the galaxies that their classification does not meet the certainty threshold. Therefore, given a certain threshold, the classifications of some galaxies do not meet that threshold, leading to an incomplete catalog. For instance, setting a threshold of 98\% might lead to a relatively accurate catalog as described in Section~\ref{evaluation}, but at the same time will not include many galaxies that belong in that morphological type but do not meet the certainty threshold. As shown by Figure~\ref{counts}, when the classification threshold is higher, the number of galaxies that their classification certainty meets the threshold gets substantially lower, and therefore many galaxies that might belong in that morphological type are left outside of the threshold cutoff.

Figure~\ref{completeness} shows the amount of galaxies that were classified by Galaxy Zoo as debiased `'superclean'', that were classified by the algorithm with certainty lower than different levels of certainty thresholds. For instance, $\sim$72\% of the galaxies classified by Galaxy Zoo as `'superclean'' spiral galaxies were not classified by the algorithm with certainty higher than 0.8, showing that improving the classification accuracy by increasing the certainty threshold leads to the sacrifice of galaxies that are of the same morphological type, but the certainty of their automatic classification does not meet the threshold.

\begin{figure}[ht]
\includegraphics[scale=0.55]{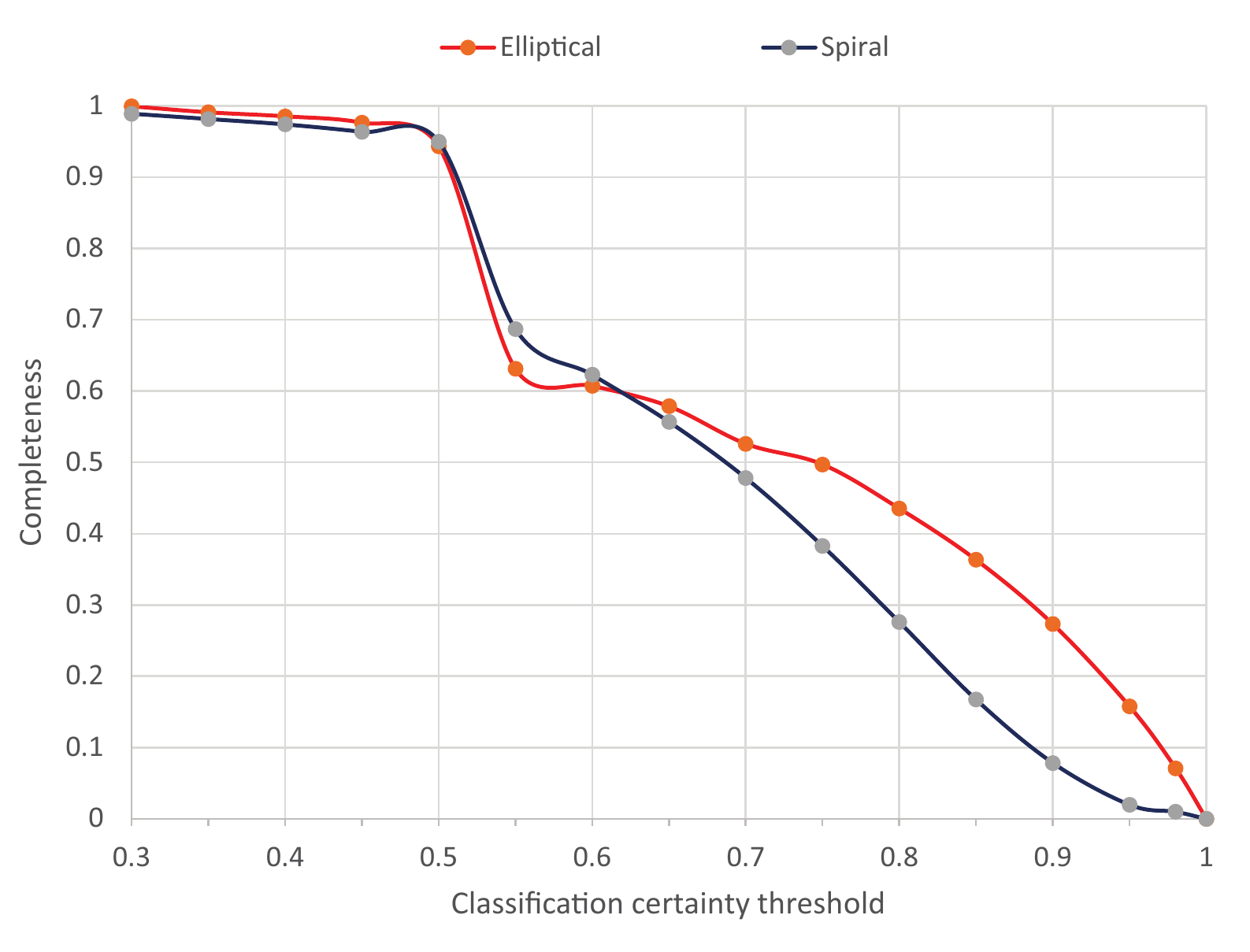}
\caption{The completeness of the catalog compared to debiased ``superclean'' Galaxy Zoo classifications.}
\label{completeness}
\end{figure}

The figure shows that when assuming all debiased ``superclean'' Galaxy Zoo galaxies are annotated correctly, the classification accuracy of $\sim$98\% of the spiral galaxies achieved when the threshold is set to 0.54 as shown in Figure~\ref{superclean_spiral_elliptical} is achieved when sacrificing $\sim$30\% of the galaxies that are actually spiral.

\begin{figure}[ht]
\includegraphics[scale=0.55]{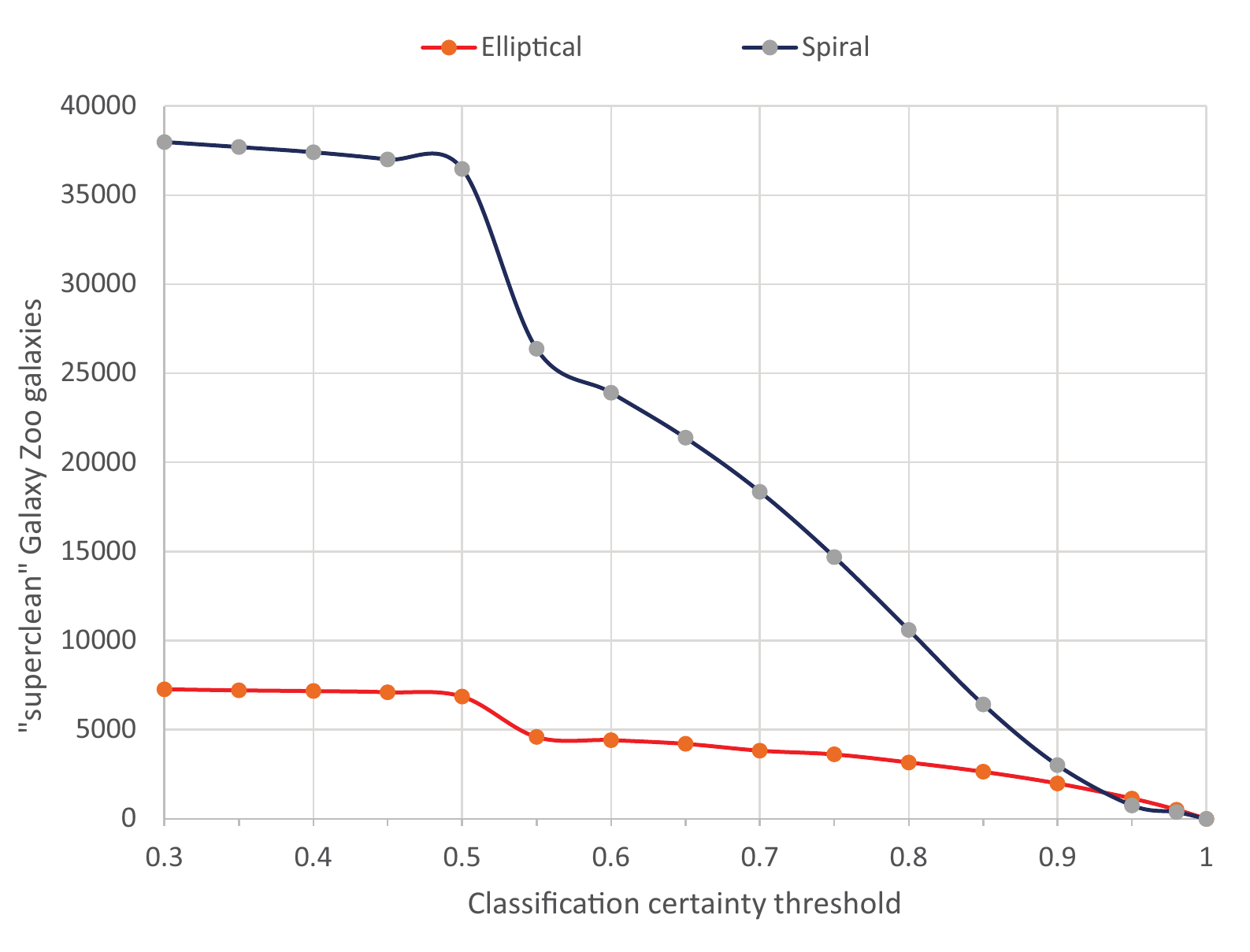}
\caption{The number of debiased ``superclean'' Galaxy Zoo classifications that were also classified by the algorithm in different certainty thresholds.}
\label{completeness_count}
\end{figure}

Figure~\ref{completeness_count} shows the number of debiased `'superclean'' Galaxy Zoo galaxies that were also classified by the automatic classifier in different certainty thresholds. Clearly, the number of `'superclean'' galaxies that were also identified by the algorithm decreases as the threshold gets higher.

Expectedly, the number of galaxies that do not meet the threshold drops sharply when the certainty threshold is below 0.5. However, a certain number of galaxies that were classified by Galaxy Zoo as ``superclean'' were misclassified by the algorithm. As Figure~\ref{completeness} shows, when the certainty is below 0.3 the number of ``superclean'' galaxies that their classification disagrees with the classification of the algorithm is very low, but it is not zero. Four debiased ``superclean'' elliptical galaxies were classified by the algorithm as elliptical with certainty of 0.3 or lower, and 420 debiased ``superclean'' spiral galaxies were classified by the algorithm as spiral with certainty lower than 0.3. Table~\ref{elliptical_missclassifications} shows the  galaxies that were classified by Galaxy Zoo as ``superclean'' elliptical but the algorithm classified them as spiral galaxies with certainty higher than 0.7 (and therefore as elliptical with certainty lower than 0.3).

\begin{table}[ht]
\begin{center}
\caption{Galaxies classified as debiased ``superclean'' elliptical by Galaxy Zoo but as spiral by the automatic classifier}
\label{elliptical_missclassifications}
\begin{tabular}{cll}
\tableline\tableline
SDSS    & Computed elliptical & Image \\
DR8 ID  & certainty   &           \\
\hline
1237650372096688281 & 0.069772 & \includegraphics[angle=00,scale=.50]{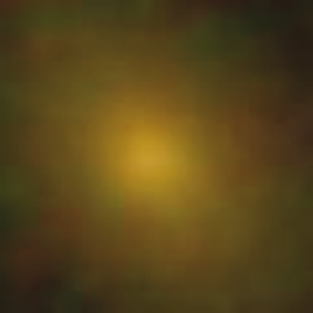} \\
1237663789575438589 & 0.092223 & \includegraphics[angle=00,scale=.50]{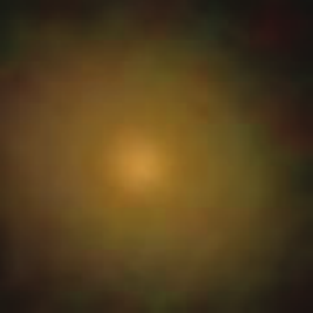} \\
1237664093439328427 & 0.093762 & \includegraphics[angle=00,scale=.50]{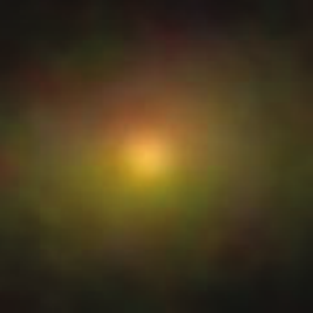} \\
1237671260125200613 & 0.084093 & \includegraphics[angle=00,scale=.50]{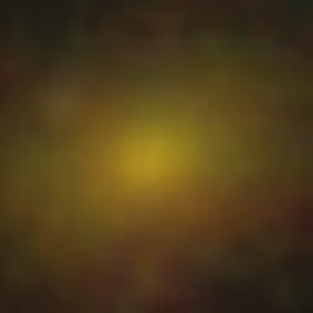} \\
\tableline
\tableline
\end{tabular}
\end{center}
\end{table}

None of these galaxies has visible spiral arms, and therefore the misclassification can be attributed to noise in the pattern recognition systems. Since machine vision image recognition systems are often not fully accurate, it can be expected that some galaxies can be misclassified, as also discussed in Section~\ref{evaluation}. All four galaxies have a clear nucleus that might also lead to higher similarity to spiral galaxies, but the galaxies in Table~\ref{elliptical_missclassifications} show that in some (rare) cases elliptical galaxies can be misclassified as spirals with relatively high certainty.

Table~\ref{spiral_missclassifications} shows example galaxies that were classified automatically as elliptical with certainty higher than 0.7, but classified as ``superclean'' spiral galaxies by Galaxy Zoo.

\begin{table}[h!]
\begin{center}
\caption{Galaxies classified as debiased ``superclean'' spiral by Galaxy Zoo but as elliptical galaxies by the algorithm}
\label{spiral_missclassifications}
\begin{tabular}{cll}
\tableline\tableline
SDSS    & Computed  & Image \\
DR8 ID  & spiral    &           \\
            & certainty   &           \\
\hline
1237651192424366347 & 0.142726 & \includegraphics[angle=00,scale=.50]{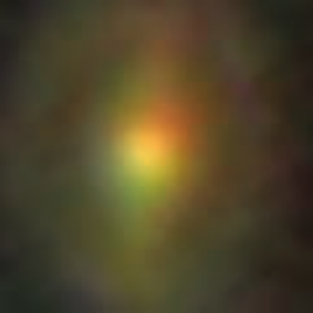} \\
1237651192429412678 & 0.290797 & \includegraphics[angle=00,scale=.50]{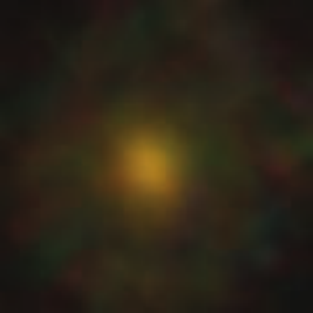} \\
1237648721759371433 & 0.190888 & \includegraphics[angle=00,scale=.50]{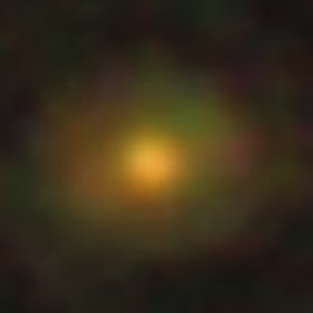} \\
1237658203978203292 & 0.120782 & \includegraphics[angle=00,scale=.50]{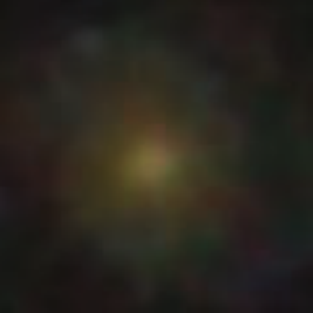} \\
1237657219874619557 & 0.241804 & \includegraphics[angle=00,scale=.50]{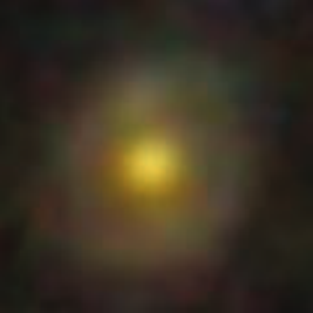} \\
1237654610682773622 & 0.189278 & \includegraphics[angle=00,scale=.50]{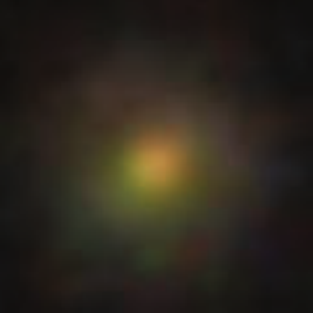} \\
1237654604795674718 & 0.272735 & \includegraphics[angle=00,scale=.50]{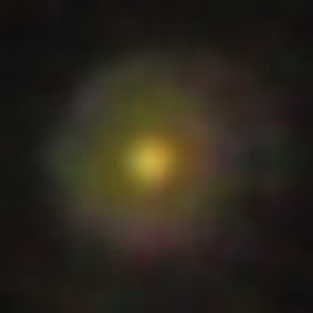} \\
1237651737381306386 & 0.115425 & \includegraphics[angle=00,scale=.50]{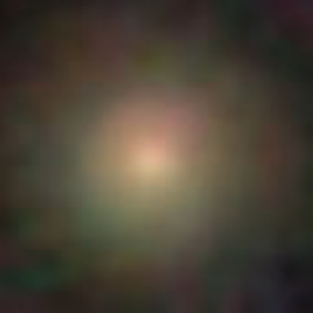} \\

\tableline
\tableline
\end{tabular}
\end{center}
\end{table}

Galaxy ID 1237654610682773622 does not show clear spiral arms, demonstrating that galaxies classified by Galaxy Zoo as ``superclean'' might also contain misclassified galaxies. The other galaxies are clearly not elliptical, and provide examples of galaxies that were misclassified by the algorithm. These galaxies normally have dimmer arms compared to their relatively bright nucleus. As discussed in Section~\ref{evaluation}, the algorithm is not in perfect agreement with manual analysis, but galaxies that are misclassified despite being classified with a high certainty are relatively rare.

\begin{figure*}[ht]
\includegraphics[scale=0.45]{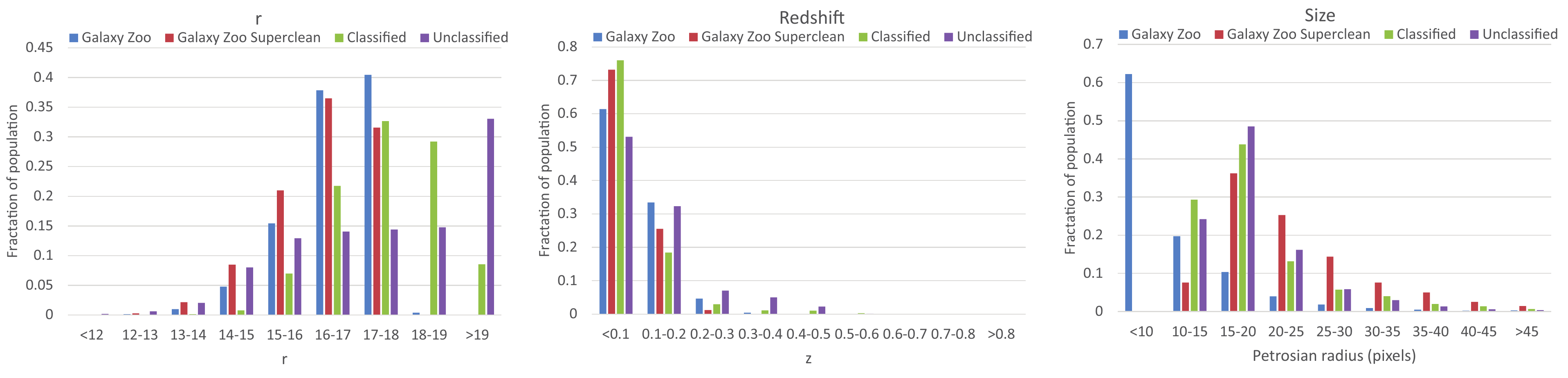}
\caption{Distribution of the r magnitude, Petrosian radius (r band), and redshift among Galaxy Zoo galaxies classified as debiased ``superclean'', galaxies of the catalog that were classified at the accuracy level of 98\% agreement with the debiased ``superclean'' Galaxy Zoo galaxies, and galaxies that their classification did not meet that level.}
\label{unclassified_histogram}
\end{figure*}

Figure~\ref{unclassified_histogram} shows the distribution of the galaxies in different ranges of r magnitude, Petrosian radius (measured on the r band) and redshift. The figure shows the distribution of Galaxy Zoo galaxies, Galaxy Zoo ``superclean'' galaxies, galaxies that were classified with certainty of 0.8 or higher for elliptical galaxies, and 0.54 or higher for spiral galaxies, which is in 98\% agreement with Galaxy Zoo debiased ``superclean'' catalog as described in Section~\ref{evaluation}, and galaxies that their classification certainty did not meet that criteria.

Expectedly, ``superclean'' galaxies are more prevalent in lower redshift, lower r, and larger galaxy size. The computer classification is also more accurate for brighter objects, and when the redshift is lower. Brighter objects provide more visual details that can be analyzed by the computer, and the correlation between r and z leads to difference in z distribution between classified and unclassified galaxies. However, the computer classification is less sensitive to size, probably because the galaxies were initially selected by their Petrosian radius and then downscaled to fit the frame as described in Section~\ref{methods}, providing a dataset that does not contain small object. 

Galaxies with higher redshift are expected to be smaller and fainter, and therefore their morphology might be more difficult to classify. Since the galaxies are selected by their size and analyzed by their morphology, the catalog can also be used to identify galaxies that SDSS spectroscopy pipeline assigned with high redshift, and have identifiable visual spiral morphology. For instance, 161 objects were classified as spiral galaxies with certainty higher than 0.54, and were associated in SDSS DR8 with spectroscopic objects with z higher than 0.4. Table~\ref{high_redshift} shows examples of these galaxies.

\begin{table}[h]
\begin{center}
\caption{Example galaxies with computed spiral certainty higher than 0.54, and were associated with spectroscopic objects in SDSS DR8 with redshift higher than 0.4.}
\label{high_redshift}
\begin{tabular}{clll}
\tableline\tableline
Number & SDSS    & SDSS & Image \\
             &   DR8 ID  & DR8 z   &           \\
\hline
1 & 1237663278465810600 & 0.889 & \includegraphics[angle=00,scale=.54]{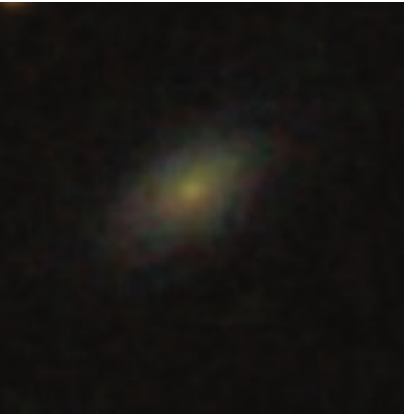} \\
2 & 1237668602608615569 & 0.428 & \includegraphics[angle=00,scale=.50]{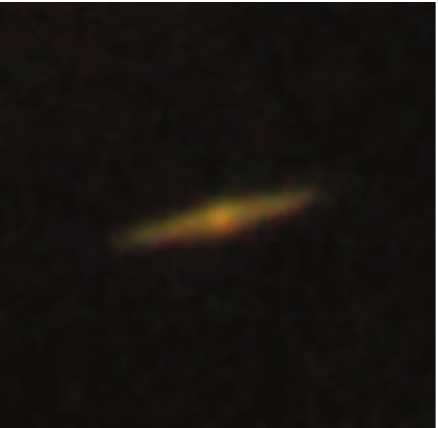} \\
3 & 1237671261743022362 & 1.044 & \includegraphics[angle=00,scale=.54]{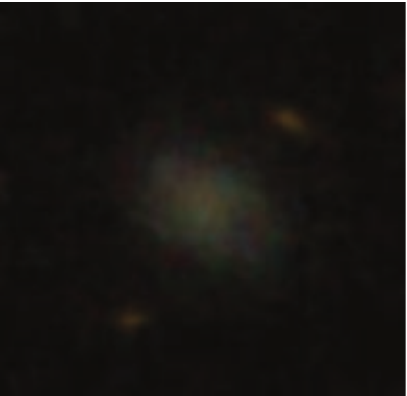} \\
4 & 1237651272422850916 & 0.534 & \includegraphics[angle=00,scale=.57]{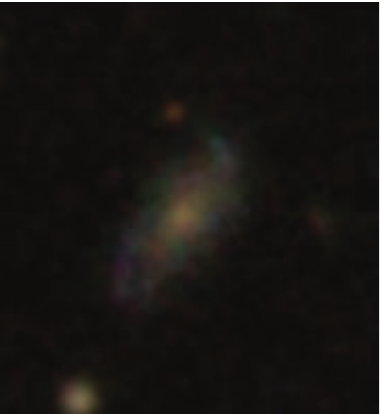} \\
5 & 1237648705126400194 & 0.908 & \includegraphics[angle=00,scale=.63]{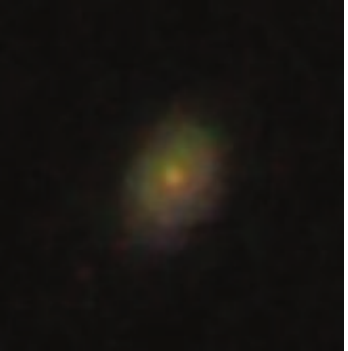} \\
6 & 1237668297140404560 & 1.262 & \includegraphics[angle=00,scale=.65]{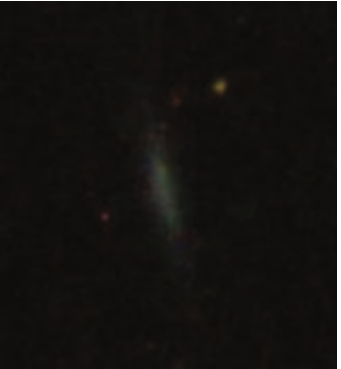} \\
7 & 1237665442603926125 & 0.998 & \includegraphics[angle=00,scale=.61]{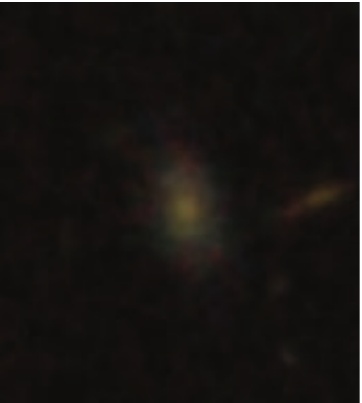} \\
8 & 1237662530608562195 & 0.401 & \includegraphics[angle=00,scale=.75]{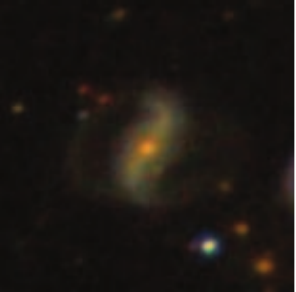} \\
\tableline
\tableline
\end{tabular}
\end{center}
\end{table}

\section{Computational complexity}
\label{computational_complexity}

As mentioned above, downloading the images of the catalog of objects without spectra from CAS was completed within $\sim$25 days. The analysis of the galaxies is done in two stages. The first is computing the numerical image content descriptors of the galaxy, and the second is the classification of the galaxy based on its numerical image content descriptors. 

Computing the numerical image content descriptors of one galaxy using a single Intel Core-i7 processor takes $\sim$45 seconds. Wndchrm can be easily parallelized so that each different instances running on different cores process different galaxies \citep{Sha08}. Using 128 cores, the numerical image content descriptors of the galaxies were computed in $\sim$12 days. The classification of all galaxies was done using a single core, and lasted $\sim$75 hours.

\section{Conclusion}
\label{conclusion}

Here we describe a catalog of $\sim$3M galaxies classified by their broad morphological type. The objects in the catalog can be selected based on the certainty of the classification to control the quantity-quality trade-off. The results show that $\sim$1.5M galaxies are classified with a degree of certainty that matches Galaxy Zoo ``superclean'' classifications in 98\% of the cases, making the catalog by far the largest of its kind.

Due to the imperfectness of the classification algorithm, higher accuracy is achieved by sacrificing some of the galaxies that do not meet a certain threshold of classification certainty, and therefore the catalog is clearly incomplete. However, given the increasing gap between the power of astronomical data collection devices and the automatic analysis tools that can turn these data into knowledge, sacrificing some of the data can still result in very large catalogs.

That also demonstrates the ability of computer programs to create clean morphological catalogs containing millions of astronomical objects. As current and future digital sky surveys have been becoming increasingly more powerful, automatic methods for analyzing galaxies are required to analyze these data. Digital sky surveys such as the Large Synoptic Sky Survey (LSST) are expected to acquire images of billions of galaxies, reinforcing the need to automatically analyze the morphology of these objects and allow scientific discoveries using these data. Therefore, computer vision methods that can produce clean catalogs of morphological descriptors of astronomical objects will become essential to fully utilized the large databases created by digital sky surveys.

\section{Acknowledgments}

This study was supported by NSF grant IIS-1546079.

Funding for the SDSS and SDSS-II has been provided by the Alfred P. Sloan Foundation, the Participating Institutions, 
the National Science Foundation, the US Department of Energy, the National Aeronautics and Space Administration, 
the Japanese Monbukagakusho, the Max Planck Society, and the Higher Education Funding Council for England. The 
SDSS Web Site is http://www.sdss.org/. 
The SDSS is managed by the Astrophysical Research Consortium for the Participating Institutions. The Participating 
Institutions are the American Museum of Natural History, Astrophysical Institute Potsdam, University of Basel, 
University of Cambridge, Case Western Reserve University, University of Chicago, Drexel University, Fermilab, the 
Institute for Advanced Study, the Japan Participation Group, Johns Hopkins University, the Joint Institute for Nuclear 
Astrophysics, the Kavli Institute for Particle Astrophysics and Cosmology, the Korean Scientist Group, the Chinese 
Academy of Sciences (LAMOST), Los Alamos National Laboratory, the Max Planck Institute for Astronomy (MPIA), 
the Max Planck Institute for Astrophysics (MPA), New Mexico State University, Ohio State University, University 
of Pittsburgh, University of Portsmouth, Princeton University, the United States Naval Observatory and the University 
of Washington.




\bibliographystyle{apj} 
\bibliography{ms}

\clearpage



\end{document}